\pdfoutput=1
\documentclass[aps,prl,twocolumn,groupedaddress]{revtex4}
 \usepackage{graphicx}
 \usepackage{dcolumn}   

\begin{document}
\title{Flow-induced channelization in a porous medium}

\author{A. Mahadevan$^1$, L. Mahadevan$^2$}
\affiliation{$^1$Department of Earth Sciences, Boston University\\
$^2$School of Engineering and Applied Sciences, Harvard University}

\date{\today} 

\begin{abstract}
We propose a theory for erosional channelization induced by fluid flow in a saturated granular porous medium. When the local fluid flow-induced stress is larger 
than a critical threshold, grains are dislodged and carried away so that the porosity of the medium is altered by erosion. This in turn affects the local hydraulic conductivity and pressure in the 
medium and results in the growth and development of channels that preferentially conduct the flow. Our multiphase model involves a dynamical porosity field that evolves along with the volume 
fraction of the mobile and immobile grains in response to fluid flow that couples the spatiotemporal dynamics of the three phases. Numerical solutions of the resulting initial boundary value 
problem show how channels form in porous media and highlights how heterogeneity in the erosion threshold dictates the form of the patterns and thus the ability  to control them.  
\end{abstract}
\maketitle

The dynamics of fluid flow through  porous continua is relevant over many orders of magnitude in length scale with applications that range from large scale flow through fractured rock in aquifers and 
oil reservoirs  to small scale flows in natural and artificially engineered tissues and gels  \cite{Bear, Stroock}. In all these cases, flows are characterized by large variations in hydraulic conductivity of 
the medium. This heterogeneity is usually ascribed to processes associated with the process of consolidation of the porous medium via the agglomeration of grains (in geology) and cells (in biology), 
but may also arise due to channels that develop in frangible porous structures as fluid flows through them. Indeed flow-induced erosive processes  on the surface of porous media have been 
implicated in the formation of patterns on planetary \cite{Malin}, littoral \cite{Hughes}  and laboratory \cite{Schorghofer} scales that involve both unconsolidated and consolidated media 
\cite{Howard}. 

However these erosive  instabilities can also occur in the bulk of fluid saturated materials where they can lead to internal channelization via the dynamic coupling of flow and hydraulic conductivity. 
Here we address the dynamical evolution of channels via flow-induced erosion within a saturated porous medium. Our theory for the active co-evolution of the phases in a porous medium is 
qualitatively different from the single phase diffusive models for the evolution of free surfaces by aggregation and erosion \cite{Barabasi} or multiphase bulk
theories for multiple fluids and/or elastic solids interacting with each other \cite{Spiegelman} but yet combines features of both in considering the fluid-induced erosion and deposition processes acting 
in the bulk of a solid skeleton. 

\begin{figure}
\centerline{\includegraphics[width=8cm]{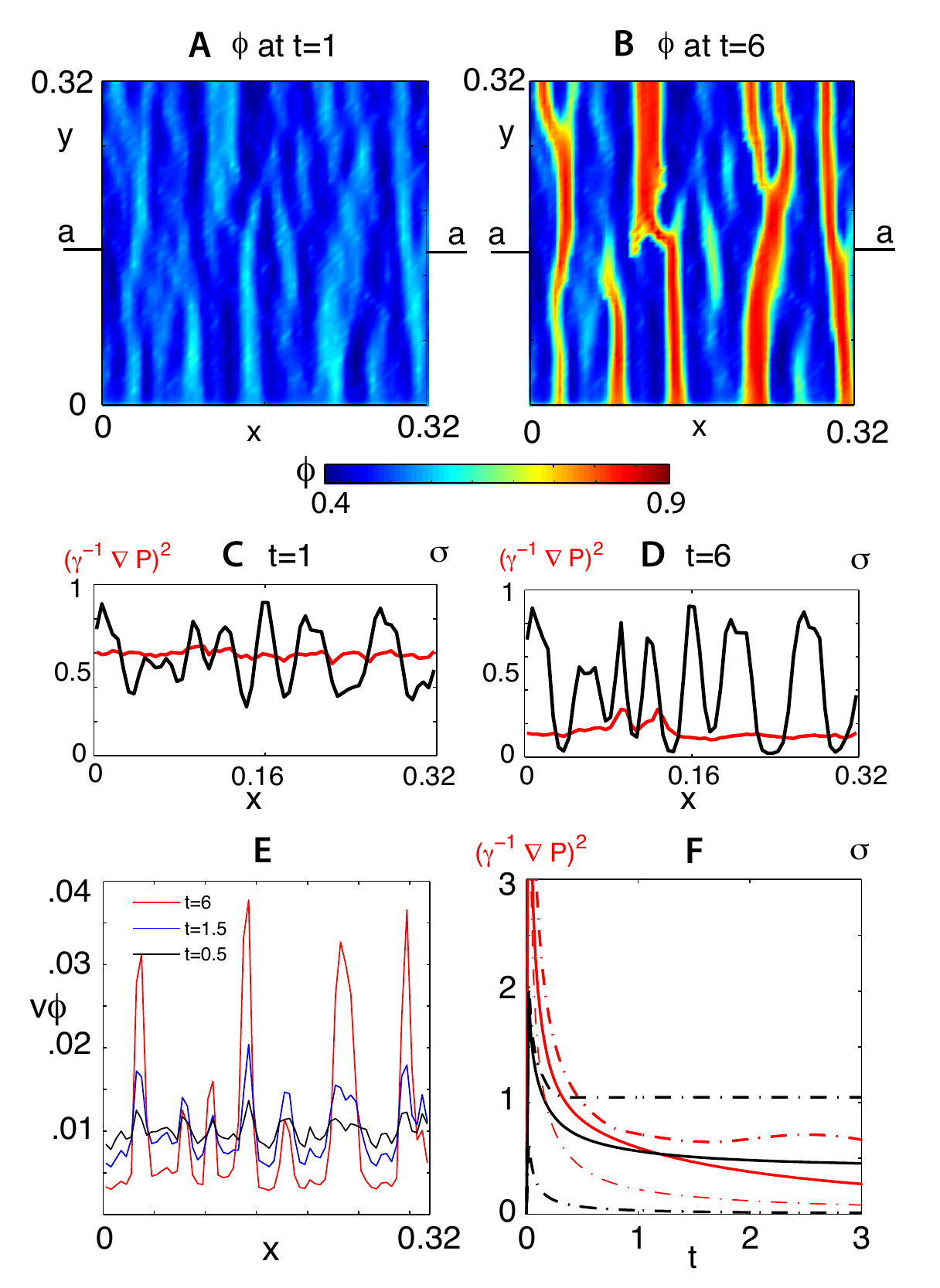}}
\caption{Numerical solution to equations (1)--(7) with a steady specific discharge $q=q_0$  prescribed at the lower boundary $y=0$, and constant pressure at the upper boundary 
$y=0.32$. (A) Spatial distribution of porosity $\phi$ at    $t=1$ and (B) at t=6. (C) $(\gamma^{-1} \nabla P)^2$ (red) and  $\sigma$ (black) are plotted along a--a (at $y=0.16$) at $t=1$ 
and (D) at $t=6$ corresponding to the panels above.  Erosion occurs where $(\gamma^{-1} \nabla P)^2 >\sigma$. At early times, this occurs at several locations. As erosion progresses, 
the pressure gradient drops, heterogeneity in $\sigma$ increases, and erosion is limited only to the channels. (E) The  flux in the $y$-direction, $v (\phi_g +\phi_l)$  plotted at section a--a 
at $t=0.5$ (black),  $t=1.5$ (red), and at $t=6$ (blue). (F) As the flux increases in eroded regions, it decreases in non-channelized regions. Here we use the $tanh$ profile for the erosion
threshold (see text), and $\Pi_1 = \Pi_2 =1$.}
\label{fig:evolve}
\end{figure}

To enable a continuum field description of the process, we consider a representative volume much larger than the grain/pore size  with $\phi_s(x,y,z,t)$ the volume fraction of the immobile solid 
phase,  $\phi_g(x,y,z,t)$, the volume fraction of the granular mobile phase,   and $\phi_l(x,y,z,t)$ , the liquid volume fraction in the medium  so that $\phi_s + \phi_g + \phi_l = 1$.  When the flow-
induced stress in a frangible porous medium exceeds a local failure threshold, particles are dislodged and mobilized \cite{Bear}. This leads to a local increase in the hydraulic 
conductivity and a decrease in the local fluid stress, even as the eroded material is carried away. Simultaneously, deposition of mobile particles can act to decrease the porosity and reroute fluid flow. 
Thus, for a given flow rate, the hydraulic conductivity evolves in space and time as a function of both the flow-induced stresses and the initial heterogeneity in the porosity of the medium, and can lead 
to complex erosional and depositional patterns. Volume conservation for the individual phases (each of which are assumed to be incompressible) implies that 
\begin{eqnarray}
\partial_t \phi_s  & =  & -e + d \\
\partial_t \phi_g & = & +e -d  - \nabla \cdot ( \phi_g  {\bf u_g} ) \\
\partial_t \phi_l=  - \partial_t ( \phi_s + \phi_g ) &  =  & -\nabla \cdot (\phi_l  {\bf u_l} )  \end{eqnarray}
where $e$ is the rate of erosion of the immobile phase, $d$ the rate of deposition of the mobile phase, and ${\bf u_g}, {\bf u_l}$ are the velocities of the granular and liquid phases, 
respectively. We note that adding equations (1)-(3) yields the global continuity equation
\begin{equation} \nabla \cdot (\phi_g {\bf  u_g} + \phi_l {\bf  u_l})=0.
\end{equation}   
We will assume that ${\bf u_g}={\bf u_l}={\bf u}$, i.e.  the granular and liquid phases have the same velocity and the effects of inertia and body forces  associated with sedimentation are 
negligible, a reasonable approximation for slow flows of nearly jammed grains. Then the continuity equation reduces to $\nabla \cdot \phi {\bf u} = 0$, where $ \phi \equiv \phi_g +\phi_l$.  
Erosion of $\phi_s$ can occur only when the fluid-induced stress exceeds a critical threshold $\sigma=\sigma (\overline{\phi_s})$, where $\overline{\phi_s} = V^{-1} \int_V \phi_s dv$.  This form of the 
threshold function characterizes the non-local nature of elastic stress distribution and failure in the porous medium in terms of a regional average of  $\phi_s$. Then we may write the local erosion rate 
$e$ as 
\begin{eqnarray}
e & = & k_e   \phi_s \left(  (\gamma^{-1} \nabla p)^2 -  \sigma(\phi_s) \right)  \ge 0, 
\end{eqnarray}
where $k_e$ is a rate, and $\gamma$ a nominal pressure gradient (based on the fluid flow rate and the hydraulic conductivity). The form of the erosion rate follows from considerations of 
symmetry: a hydrostatic pressure $p$ cannot lead to erosion, but a gradient in $p$ can. However, the sign of the gradient
is not important, so that we  have chosen the simplest analytic dependence consistent with this symmetry \footnote{Using the asymptotically correct but non-analytic form $|\nabla p|$  yields  
qualitatively similar results.}. We assume $ \sigma =0.5 (\tanh(2 \pi ( \overline{\phi}_s- 0.6) ) +1 ),  0\leq \sigma \leq 1$ to mimic the sharp dependence of the failure stress on 
the volume fraction, although later we will consider other forms as well.  A simple model for the local 
deposition rate $d$, the rate at which the mobile granular material is converted back to the immobile solid phase, is given by 
\begin{equation}
d= k_d  (\phi_s -\phi_s^*) \; \phi_g, \mbox{~~~where~~~} d\ge 0.
\end{equation}
The form of the deposition with a rate $k_d$  is based on a binary collision picture -- mobile grains will come to rest only if they interact  with immobile grains, with a threshold $\phi_s^*$.  
In the porous medium, we assume that the volumetric flow rate per unit cross-sectional area, i.e. the specific discharge  ${\bf q} \equiv {\bf u} ( \phi_l + \phi_g) $,  is 
given by Darcy's law\footnote{This can be generalized to a Brinkman-like equation if necessary in regions of high porosity.} 
\begin{equation}
{\bf q} \equiv {\bf u} \phi = - D(\phi) \nabla p, \mbox{~where~} D =\frac{ \phi^3 \: l_g^2}{A \mu \: (1 - \phi^2  )}.\label{eq:darcy} 
\end{equation}
Here, $D(\phi)$, the hydraulic conductivity, is assumed to follow the Carman-Kozeny relation \cite{Bear} and is in general a nonlinear function of the local fluid (pore) volume fraction $\phi = \phi_l + \phi_g$.  Here $l_g$ is the nominal pore size (which scales with the grain diameter), $\mu$ is the dynamic viscosity of the interstitial fluid.
Using $l_g \sim 1mm$ and the dimensionless constant $A \sim 10^2$ for spherical grains \cite{Bear}  
results in $D \sim 10^{-5} m^3s/kg $ in a water saturated medium, which we assume to be isotropic. Erosion leads to a wide range in $\phi$, and consequently, variations in $D$.

\begin{figure}
\centerline{ \includegraphics[width=8cm]{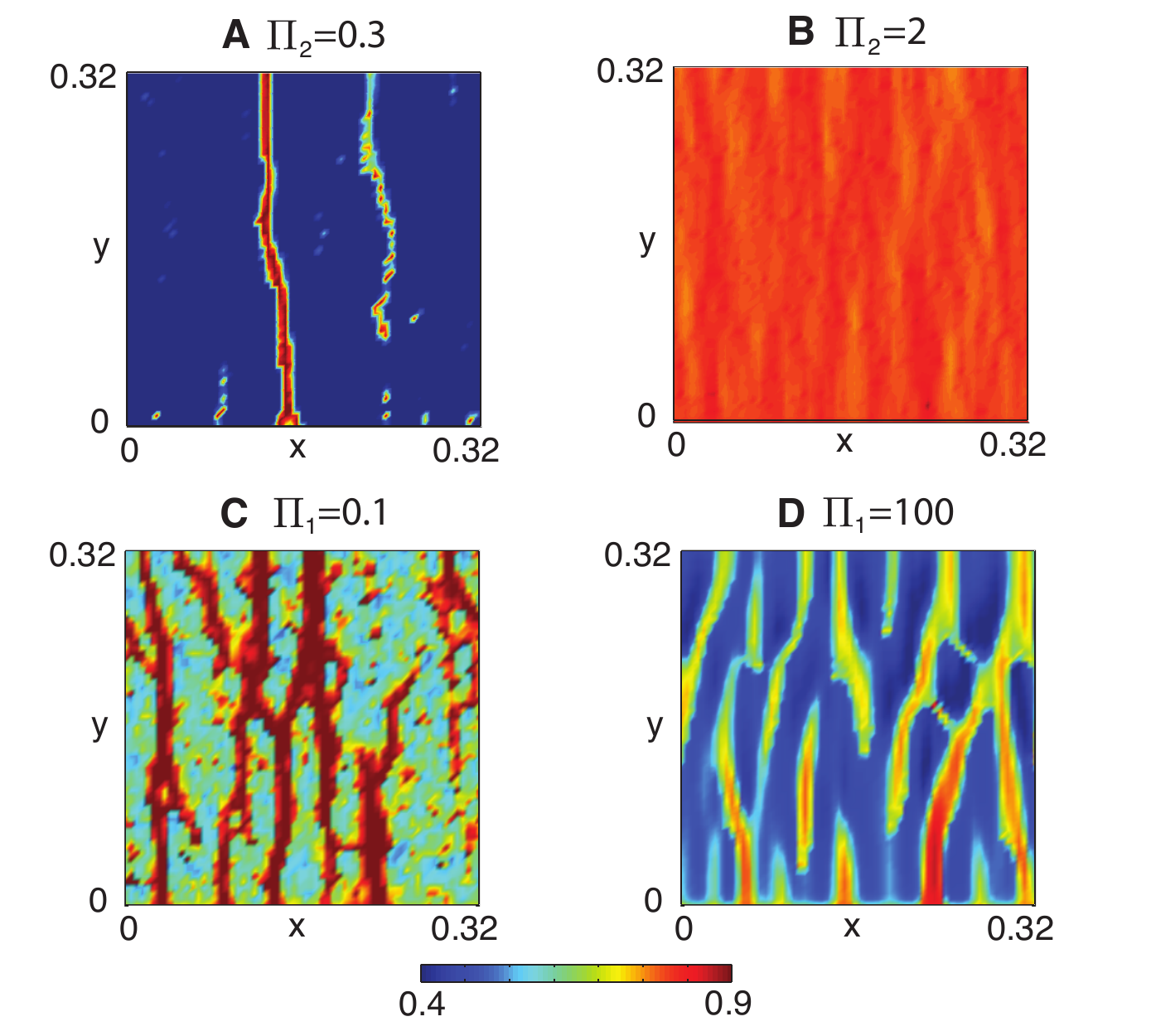}}
\vspace{-0.4cm}
\caption{Result of varying the model parameters on the final distribution of porosity $\phi$ plotted at $t=6$. The scaled specific discharge $q$ specified at $y=0$  is  (A) decreased to $0.3$ and (B) increased to $2$  Large values of $q$ lead to complete erosion, whereas weak $q$ leads to no erosion. (C) The erosion rate $k_e$ is increased by a factor of 10, i.e. $\Pi_1= 0.1$, and (D) deposition rate $k_d$ is increased by a factor of 100, i.e. $\Pi_1= 100$.}
\end{figure}

We use the domain size, $L=1m$, specific discharge $q_0=1 cm/s$, and time $T=L/q_0$ \cite{Kudrolli} to make the problem dimensionless, so that the dimensionless parameters in the problem 
include the thresholds for 
erosion and deposition $\sigma$ and $\phi_s^*$,  as well as  the ratio of deposition to erosion rates $\Pi_1= \frac{k_d}{k_e}$, and  the ratio of advection to erosion rate $ \Pi_2=\frac{q_0}{k_e L}
$.  We solve equations (1--7) numerically in 2-dimensions ($x,y$) using a finite volume method, with a constant scaled specific discharge ${\bf q}=(0,q)$ at the inflow boundary $y=0$, while at the outflow 
boundary  $y=L_y$ we set pressure $p= 0$ (atmospheric pressure). In the $x$ direction, we use periodic boundary conditions at $x=0$ and $x=L_x$, with a square domain of dimension $L_x=0.32, 
L_y= 0.32$ and a uniform grid resolution $\Delta = 0.05$. Prescribing the inlet pressure instead of the inlet discharge leads to either no erosion (if $(\gamma^{-1} \nabla p)^2 <\sigma$ ) or catastrophic 
erosion if the pressure gradients are larger than the threshold. Thus we prescribe a specific discharge at the inlet,  and a vanishing pressure at the outlet. The pressure is determined by iteratively 
solving the Poisson equation $\nabla  (D(\phi) \nabla p)=0$ obtained by substituting (7) into (4), then calculating the erosion rate $e$ and the deposition rate $d$ and finally evolving equations (1)--(3) 
to update the volume fraction of the phases $\phi_s, \phi_g,  \phi_l$ from one time step to the next, with time step  $\Delta t=10^{-4}$. Starting with an initial mean volume fraction of mobile grains $
\phi_g=0$ throughout the domain and a mean liquid volume fraction $\phi_l=0.2$ with an additive white Gaussian noise (standard deviation $sd=\sqrt{\langle \phi_l^2 \rangle - \langle \phi_l \rangle^2}^{1/2} =0.01$) which leads to 
variations in $\sigma$ in the medium, we allow the system to evolve until it reaches a quasi-steady state. The non-local form of the erosion threshold $\sigma(\overline{\phi}_s)$, where the overbar 
denotes a weighted spatial average is calculated numerically as $\overline{\phi}_s(i,j)=0.2 \phi_s(i,j) + 0.1 \phi_s(i\pm1, j \pm 1)$ (thus averaging over the 8 surrounding neighbors of a grid point) and 
amounts to a radius of influence of the fluid-induced stress that extends a few grain diameters.

\begin{figure}
\centerline{ \includegraphics[width=8cm]{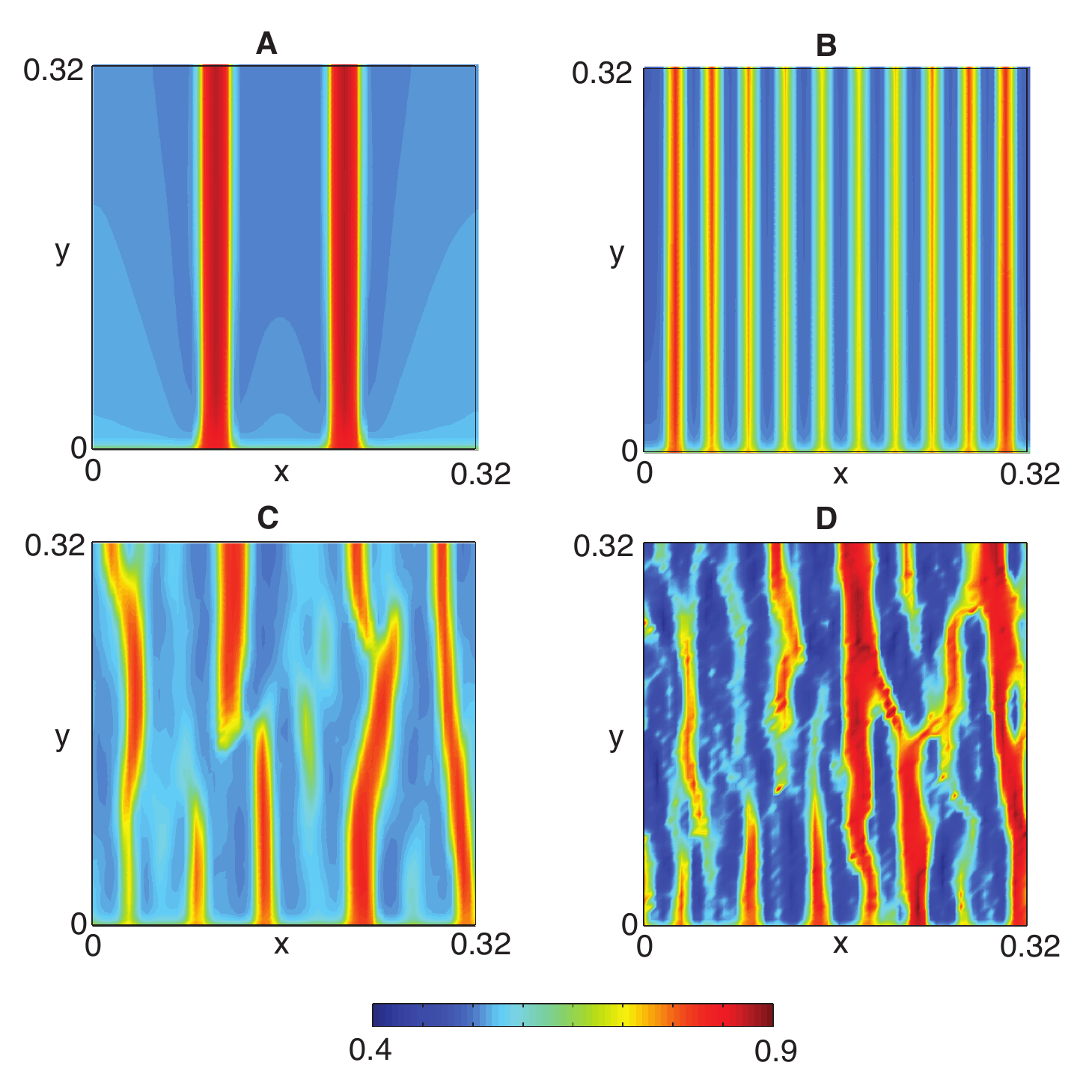}}
\vspace{-0.5cm}
\caption{The final distribution of porosity (at $t=6$) is shown to be sensitive to the initial heterogeneity in $\sigma$ arising from $\phi_s$.  The upper row shows the porosity distribution resulting when 
the initial distribution of $\phi_s$ is set to 0.8  everywhere, except in specific places where it is decreased by 1\% from the uniform background value (A) along two lines, and (B) along 10 lines,  each 
being one grid cell wide.  In the lower row, the standard deviation ($sd$) in the initial perturbation to $\phi_s$ is varied from its previous  value of $sd=0.01$. (C) $sd= 0.001$. (D) $sd=0.03$. }
\end{figure}

In Fig.~1A,B  we show two snapshots of a porous medium being evolved with $q=0$ at the inlet boundary, and $\Pi_1=\Pi_2=1$, as it starts to erode inhomogeneously and reaches its final channelized state. Dynamically, this process arises via positive 
feedback: flow is enhanced through the regions of low solid fraction (high hydraulic conductivity) as the fluid scours out a channel, while regions with a higher solid fraction (and strength) and lower 
hydraulic conductivity are circumvented by the flow. Indeed, in Fig.~1C,D we see the interplay between the heterogeneity in the erosion threshold $\sigma$ and the squared pressure gradient $
(\gamma^{-1} \nabla p)^2$, leading to an enhanced flow through regions of high hydraulic conductivity at the expense of low flow through other regions with the passage of time (Fig.~1E) even as the 
total flow rate remains the same. In Fig. 1F, we provide a global view of the process: erosion continues until the average porosity over the entire domain increases sufficiently so that the pressure 
gradient everywhere falls below the threshold for erosion and the system approaches a quasi-steady state, where erosion and deposition become vanishingly small everywhere. 

To understand how this steady state depends on the dimensionless parameters, we first vary  the scaled specific discharge $q$ at the inlet. When $q < q_c$, i.e. $\Pi_2<\Pi_c$, a critical
scaled flow rate that depends on the initial porosity distribution in the medium, no erosion or channelization occurs, because the pressure gradients everywhere are smaller than the erosion threshold. For  $
\Pi_2=\Pi_c$, a single narrow channel and secondary partial channel are formed as shown in Fig.~2A. As $\Pi_2$ is increased further, the number of channels as well as the width of channels 
increases (Fig.~1B); for even higher flow rates, the entire medium begins to erode away as shown in Fig.~2B. In all cases, the mean steady-state porosity increases with the specific discharge, linearly 
at first, before it asymptotes slowly to a steady state. Varying the erosion and deposition rates also leads to variations in the erosion patterns; in Fig.~2C, we show the erosional pattern resulting from a 10-fold increase in 
$k_e$ ($\Pi_1=0.1$), which leads to the faster evolution of channels. Conversely, increasing $k_d$ 100-fold so that $\Pi_1=100$ increases deposition and leading to the blockage of channels (Fig.~2D), although the 
average number or size of channels does not change in either case relative to when $\Pi_1=1$ (corresponding to Fig.~1B). 

\begin{figure}
\centerline{ \includegraphics[width=8cm]{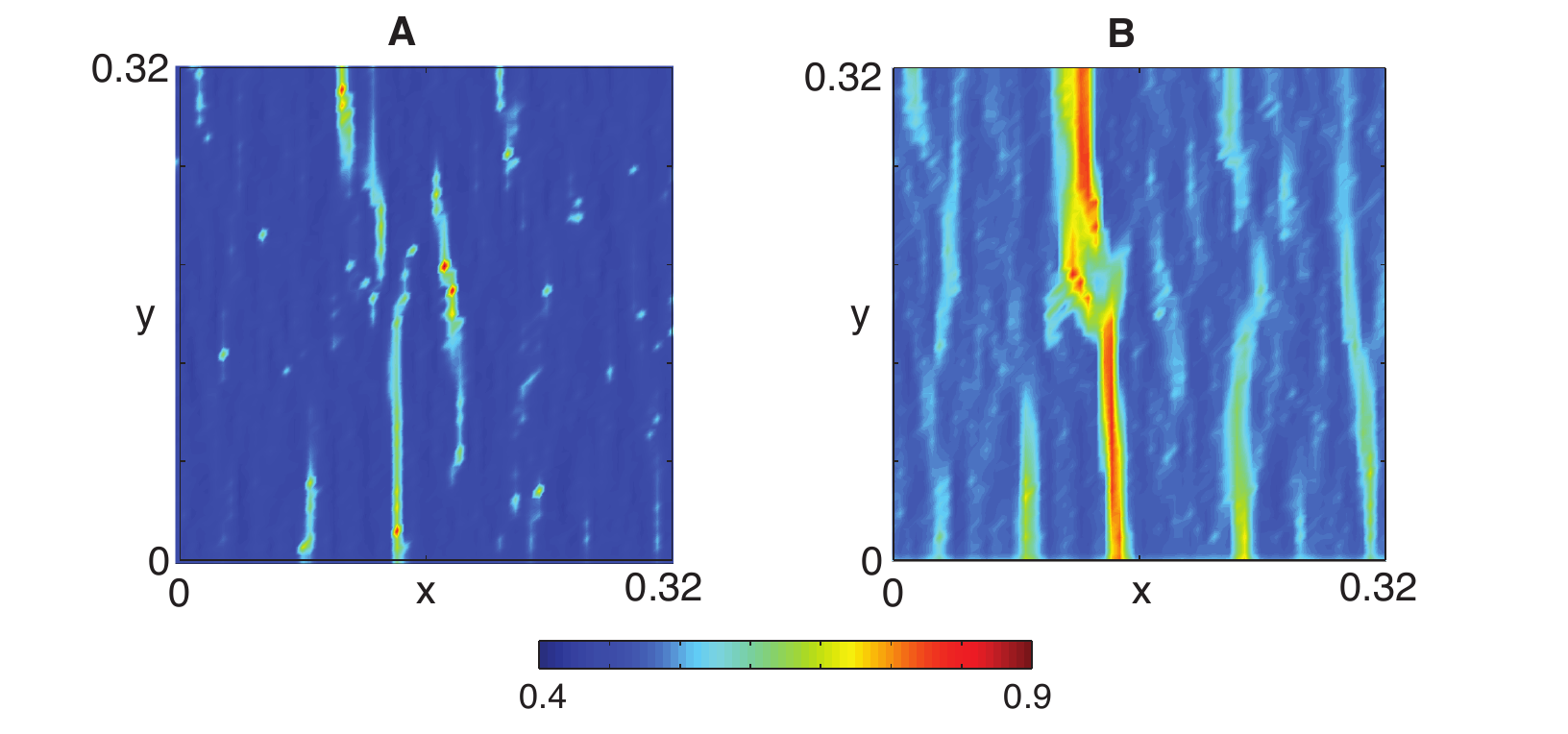}}
\vspace{-0.4cm}
\caption{The evolution of the porosity is sensitive to the functional form of the erosion threshold $\sigma$. Here the final distribution of porosity $\phi$ is shown for other choices of $\sigma$.   (A) $
\sigma = \overline{\phi}_s$, (B) $\sigma = \overline{\phi}_s^2$.  These results can be compared with Fig.~1, where $\sigma = 0.5 (\tanh(2 \pi ( \overline{\phi}_s- 0.6) ) +1 ) $ (where subtracting 0.6 instead of 
0.5 provides a slight asymmetry to the $\tanh$ profile.)}
\end{figure}

A question that naturally arises is the mechanism for the selection of channel spacing and width when fluid flows through a nominally homogeneous porous medium. The natural length scales in 
the problem are the system size $L$ the nominal pore size $l_p$ which evolves with time, but remains a microscopic length, and the length scales $q_0/k_e, q_0/k_d$; the latter control the
dynamical evolution of the channels but not their final state.  Increasing the domain size does not affect the number or size of channels. What remains is the threshold for erosion $\sigma$; since the onset of channelization is strongly influenced by fluctuations in the 
porosity (and thus the fragility) of the medium, we expect that linear stability analysis of the base state should predict that channels form at locations where $\sigma$ is smallest  initially. Thus for the
same inlet specific discharge, the size and number of channels is a function of $\sigma(x,y,0)$. In Fig.~3A we show that for a given inlet specific discharge, if $\sigma(x,y,0)\equiv f(x)$ has a single 
minimum, a single channel forms and grows until the pressure gradient falls below the erosion threshold, while in Fig.~3B, we see that if $\sigma(x,y,0)$ has multiple minima, multiple channels 
form. Of course, it is not sufficient to consider the mean value of the threshold; instead one must account for the complete probability distribution of the erosion threshold. For our simple Gaussian
model of disorder, if the variance in the threshold for erosion (or equivalently the porosity fluctuations) is also changed, this leads to variations in the patterns as well. In Fig. ~3C,D, we show that an 
increase in the standard deviation of the initial white noise leads to greater heterogeneity in the channel number and spacing. 

Finally, we consider  the functional form of the erosion threshold $\sigma(\overline{\phi_s})$. In Fig.~4A we see that for $\sigma= \overline{\phi}_s$ the final morphology of the erosion patterns is more 
uniform compared to that shown in Fig.~4B for $\sigma= \overline{\phi}_s^2$, which is itself less variable than for the case when $\sigma$ follows a $tanh$ profile (Fig.~1-3). We thus see that
the form of the erosion threshold function, and its initial, possibly heterogeneous, spatial distribution are crucial in determining the growth and form of the channels, which a simple linear analysis alone
cannot capture.

Our theory of flow-induced channelization in porous media focuses on the simplest facets of the phenomena that involve changes in porosity,  pressure gradients and 
flow in a minimal multiphase theory that involves fluid, granular and immobile phases interacting with each other.  Below a critical flow rate, little 
or no erosion occurs. Above this threshold, the porous medium starts to erode heterogeneously at locations where the critical threshold is lowest; positive feedback then enhances erosion 
locally in other frangible regions leading to oriented regions of higher porosity that are the hallmarks of channels. 

Recent experiments  with bidisperse granular mixtures in a Hele-Shaw cell 
\cite{Kudrolli} are qualitatively consistent with the strong dependence of the erosional patterns on initial heterogeneities in the strength of the porous medium. A natural next step is to compare 
experiments with theory quantitatively, and in particular to understand the physics associated with the microscopic parameters $\sigma, k_e,k_d$ in laboratory and terrestrial physical systems
and possible generalizations to biological systems in the context of vascularization in  natural and artificial settings \cite{angiogenesis}. 

\begin{acknowledgments} 
We thank A. Kudrolli and A. Orpe for sharing their observations of erosional patterns and for discussions,  DOE NICCR (AM), and  Harvard-NSF MRSEC  (LM) for support. 
\end{acknowledgments}

\end{document}